\documentclass[conference]{IEEEtran}
\IEEEoverridecommandlockouts
\usepackage{cite}
\usepackage{amsmath,amssymb,amsfonts}
\usepackage{algorithmic}
\usepackage{graphicx}
\usepackage{dblfloatfix}
\usepackage{textcomp}
\usepackage{xcolor}
\usepackage{float}
\usepackage{comment}
\usepackage{times}  %
\usepackage{mathptmx}

\usepackage{mathptmx}
\def\BibTeX{{\rm B\kern-.05em{\sc i\kern-.025em b}\kern-.08em
    T\kern-.1667em\lower.7ex\hbox{E}\kern-.125emX}}

\begin{document}


\title{Symbol\,Rate\,Maximization\,in\,Rolling-Shutter\,OCC: Design and Implementation Considerations \vspace{-2mm}}

\author{\IEEEauthorblockN{Xinyu Zhang\IEEEauthorrefmark{1}, Alexis A. Dowhuszko\IEEEauthorrefmark{1}, Miguel Rêgo\IEEEauthorrefmark{2}, Pedro Fonseca\IEEEauthorrefmark{2}, Luís Nero Alves\IEEEauthorrefmark{2}, \\Jyri H\"am\"al\"ainen\IEEEauthorrefmark{1}, Risto Wichman \IEEEauthorrefmark{1}
 } 
\vspace{1mm}
\IEEEauthorblockA{
\IEEEauthorrefmark{1}Department of Information and Communications Engineering, Aalto University, Finland. \\
\IEEEauthorrefmark{2}Instituto de Telecomunicações, Universidade de Aveiro, Aveiro, Portugal .
}
 \vspace{2mm}
\IEEEauthorblockA{Emails: \{xinyu.1.zhang, alexis.dowhuszko, jyri.hamalainen, risto.wichman\}@aalto.fi; \{miguel.rego, pf, nero\}@ua.pt}
\vspace{-6mm}
}
\maketitle
\begin{abstract}
Optical Camera Communication~(OCC) systems can take advantage of the row-by-row scanning process of rolling-shutter cameras to capture the fast variations of light intensity coming from Visible Light Communication~(VLC) LED-based transmitters. In order to study the maximum data rate that is feasible in such kind of OCC systems, this paper presents its equivalent digital communication system model in which the rolling-shutter camera is modeled as a rectangular matched-filter whose time width is equal to the exposure time of the camera, followed by a sampling process at the pixel row sweep rate of the camera. Based on the proposed rolling-shutter camera model, the maximum symbol rate that such OCC systems can support is experimentally demonstrated, and the impact of imperfect time synchronization between the VLC transmitter and the rolling-shutter OCC receiver is characterized in the form of Inter-Symbol Interference~(ISI). The equivalent three-tap channel model that results from this process is experimentally validated and the generated ISI is compensated with the use of linear equalization in reception. Simulation and experimental results show a strong correlation between them, demonstrating that the proposed approach can be used to make the OCC system work at the Nyquist sampling rate, which is equivalent to the pixel row sweep rate of the rolling-shutter camera used in reception.
\end{abstract}

\vspace{2mm}


\section{Introduction}
Visible light communication~(VLC) is a promising optical wireless communication technology that employs intensity modulation to reuse existing illumination infrastructure for data communication services~\cite{chow2025optical,nguyen2025survey}. Different Light-Emitting Diode~(LED) technologies are typically employed to transmit multi-level data symbols, such as \mbox{$M$-ary} Pulse Amplitude Modulation ($M$-PAM), in which the information is conveyed by varying the intensity of the light pulses that are emitted. The most commonly used light detectors in VLC are black silicon Photodiodes~(PDs), which convert the incident optical power into an electrical signal that is subsequently processed for symbol detection. However, most commercial user terminals available on the market today do not include an embedded PDs that could be used to receive fast-varying light signals, hindering the widespread adoption of VLC technology in our daily applications. Fortunately, smartphones, tablets, and similar devices come with cameras that are suitable for Optical Camera Communication~(OCC)~\cite{saeed2019optical}, serving as a transition step towards the massive adoption of VLC technology.

There are two predominant types of methods used in digital cameras to capture images, particularly in those using CMOS sensors: global-shutter cameras, which simultaneously expose all pixels in the captured image, and rolling-shutter cameras, which sequentially expose and read out individual rows within the image, offering the possibility to capture fast intensity variations on the light emitted by an LED. Then, assuming that the LED symbol time and the row sampling time of the camera are identical, and that perfect time synchronization exists between the LED transmitter and the OCC receiver, each row of the image captured by the camera will accurately register a single intensity-modulated symbol~\cite{matsunaga2023exposure,matsunaga2022q,higa2024demonstration}. However, such time alignment is not possible in practice because the cameras of user terminals cannot synchronize their row sweep time to the clock signal of the VLC transmitter~\cite{liu2020some}. As a result, most practical OCC implementations reduce the symbol rate to enable a row-wise Over-Sampling Factor~(OSF) in reception, such that each VLC symbol is registered by few rows of the rolling-shutter camera\cite{rego2025experimental}. In this situation, image rows near VLC symbol edges are usually interfered by the light intensity from adjacent symbols, and the value of the pixels in the central row of the horizontal stripes that registers the LED blinking is used to detect the transmitted symbol. However, this oversampling strategy in reception notably limits the maximum symbol rate that the OCC system can support.

Most smartphones and consumer cameras can typically operate at $60$ frames-per-second~(fps) with about $1080$ rows per image in full high definition~\cite{liu2020some}. In this situation, the maximum sampling rate in reception is equal to $64.8$~kilosamples-per-second~(ksps). However, when an OSF is applied in reception to deal with time synchronization issues, the resulting symbol rate is reduced to few tens of kilobits-per-second~(kbps)~\cite{shi2024nearest}. To address this challenge, extensive research efforts have been devoted to improving transmission efficiency and signal reliability. 
For example, the author of~\cite{dong2025error} established a comprehensive model covering optical emission, propagation, imaging, and pixel integration, treating the camera as an integrate-and-sample system. By defining a so-called Exposure-to-Symbol Ratio~(ESR), the authors analytically revealed the quantitative relationship that exists among exposure time, Inter-Symbol Interference~(ISI) and Bit Error Rate~(BER), providing a solid theoretical basis for performance optimization. Unfortunately, this study only focused on the general impact of ESR, and did not provide a detailed analysis of the most interesting case in which the exposure time equals the transmitted symbol time.

Concerning more experimental works, the authors of~\cite{rego2024using,rego2025experimental} developed an  OCC system using a low-cost Raspberry Pi Camera V2.1, in which the use of \mbox{$32$-PAM} significantly enhanced the achievable data rate. Although this study demonstrated that the use of high-order modulations can effectively improve the OCC throughput even with low-cost hardware, the LED data rate was limited to $10$\,kilosymbols-per-second to ensure reliable ISI-free symbol sampling. 
Similarly, a Nearest-Neighbor Bit-Aided Decision~(NNBAD) scheme was proposed in~\cite{shi2024nearest} for a low-complexity ISI suppression using an OCC receiver with few pixel rows per symbol. Finally, the authors of~ \cite{chen2024frame} proposed a Frame-Rate-Adaptive Fractionally Spaced Equalizer (FA-FSE) that dynamically compensates for timing offsets and ISI. However, these methods still rely on oversampling of the captured VLC signal in the OCC receiver.

To address this limitation, this paper starts by presenting the system model of an OCC system. After that, it proposes an equivalent OCC receiver model that accurately represents the integration process that is carried out by the rolling-shutter camera in reception, which is equivalent to the matched filtering plus symbol-sampling processing done in traditional digital communication systems. This way, the reason for using VLC symbol rates that are a few times slower than the camera row sweep rate becomes evident. However, this paper extends the analysis to the upper-bound case in which the VLC symbol duration equals the OCC exposure time, considering both perfect and non-perfect time synchronization conditions. Finally, channel estimation and linear equalization are applied in the OCC receiver to mitigate the ISI due to imperfect synchronization, enabling to recover the transmitted symbol sequence at the upper-bound symbol rate limit that the rolling-shutter camera row sweep rate supports.

The rest of this paper is organized as follows: Section~\ref{sec:2} presents the equivalent OCC system model from the perspective of digital communications. Section~\ref{sec:3} studies the effect of non-synchronous-sampling in the OCC receiver, including channel estimation and equalization techniques to mitigate the ISI that is added in the process. Section~\ref{sec:4} presents the simulation camera model and reports the numerical simulation results and experiments carried out with commercial off-the-shelf cameras. Finally, conclusions are drawn in Section~\ref{sec:5}.

\section{System Model}
\label{sec:2}

This section presents the OCC system model from the perspective of a conventional digital communication system, demonstrating that the use of rectangular pulses in the VLC transmitter is the best option to maximize the Signal-to-Noise Ratio~(SNR) of the signal samples captured by the pixel rows of the rolling-shutter camera in reception.

\subsection{Rolling-shutter OCC system model}
\label{sec:2a}
\begin{figure}[t]
\centering
\includegraphics[width=0.48\textwidth]{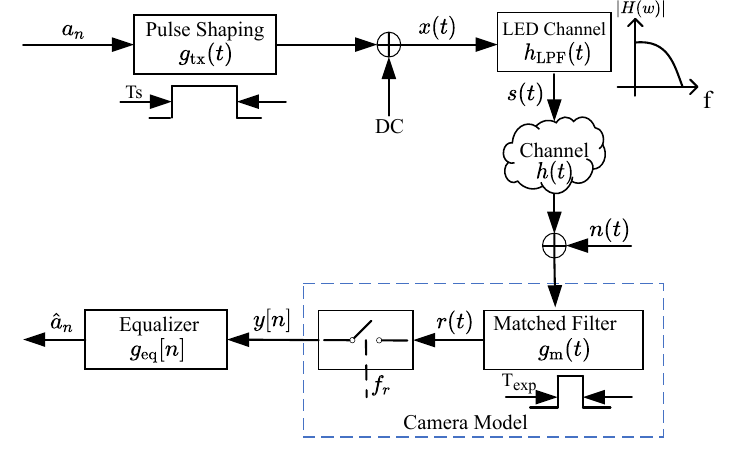}
\vspace{-2mm}
\caption{System model of the proposed OCC system. The VLC rectangular pulses have a duration $T_{\rm s}$, whereas the camera exposure time is equal to $T_{\rm exp}$.}
\label{fig_1}
\vspace{-2mm}
\end{figure}

The proposed OCC system consists of three main blocks, namely the VLC transmitter, the optical wireless channel, and the rolling-shutter camera-based receiver followed by a linear equalizer, as illustrated in~Fig.~\ref{fig_1}. When $M$-PAM modulation is used in transmission, the VLC symbols can be expressed as
\begin{equation}
a_n \in \{\pm 1, \pm3,...,\pm(M-1)\}/(M-1),
\label{eq:2.1}
\end{equation}
where $M = 2^b$ to accommodate $b$ bits per $M$-PAM symbol. The normalization in~\eqref{eq:2.1} ensures that the maximum intensity does not exceed the peak power that the LED can transmit.

To form the transmitted waveform, a rectangular pulse  
\begin{equation}
\label{eq:2}
g_{\text{tx}}(t)=
\begin{cases}
I_{0}, & 0 \leq t \leq T_s,\\[4pt]
0, & \text{otherwise},
\end{cases}
\end{equation}
is used, where $T_s$ denotes the symbol duration and $I_0$ is the nominal optical intensity of the light source. To ensure that the entire range of intensity levels to be transmitted are positive, a DC bias is added such that the lowest (highest) intensity level corresponds to a non-zero (the maximum LED) light intensity value. Thus, the continuous time-domain baseband signal that is transmitted can be written as
\begin{equation}
x(t) = \sum_n a_n \,  g_{\text{tx}}(t - n \, T_s) + I_0,
\label{eq:2.3}
\end{equation}
where $n$ represents the $M$-PAM symbol index in the transmitted symbol sequence and $I_0 = I_{\max}/2$. Finally, this equivalent unipolar signal is then used to drive the light intensity that is emitted by the LED, which can be represented as 
\begin{equation}
s(t) = x(t) * h_{\text{LPF}}(t)
\label{eq:2.4}
\end{equation}
where $h_{\rm LPF}(t)$ models the low-pass response of the LED and $f(t) * g(t) = \int_{-\infty}^{\infty} f(\tau) \cdot g(t - \tau) \, d\tau $ is the convolution operation. The transmitted signal is affected by geometric loss and Additive White Gaussian Noise~(AWGN) when propagates through the optical wireless channel.

\begin{figure*}
\centering
\includegraphics[width=0.95\textwidth]{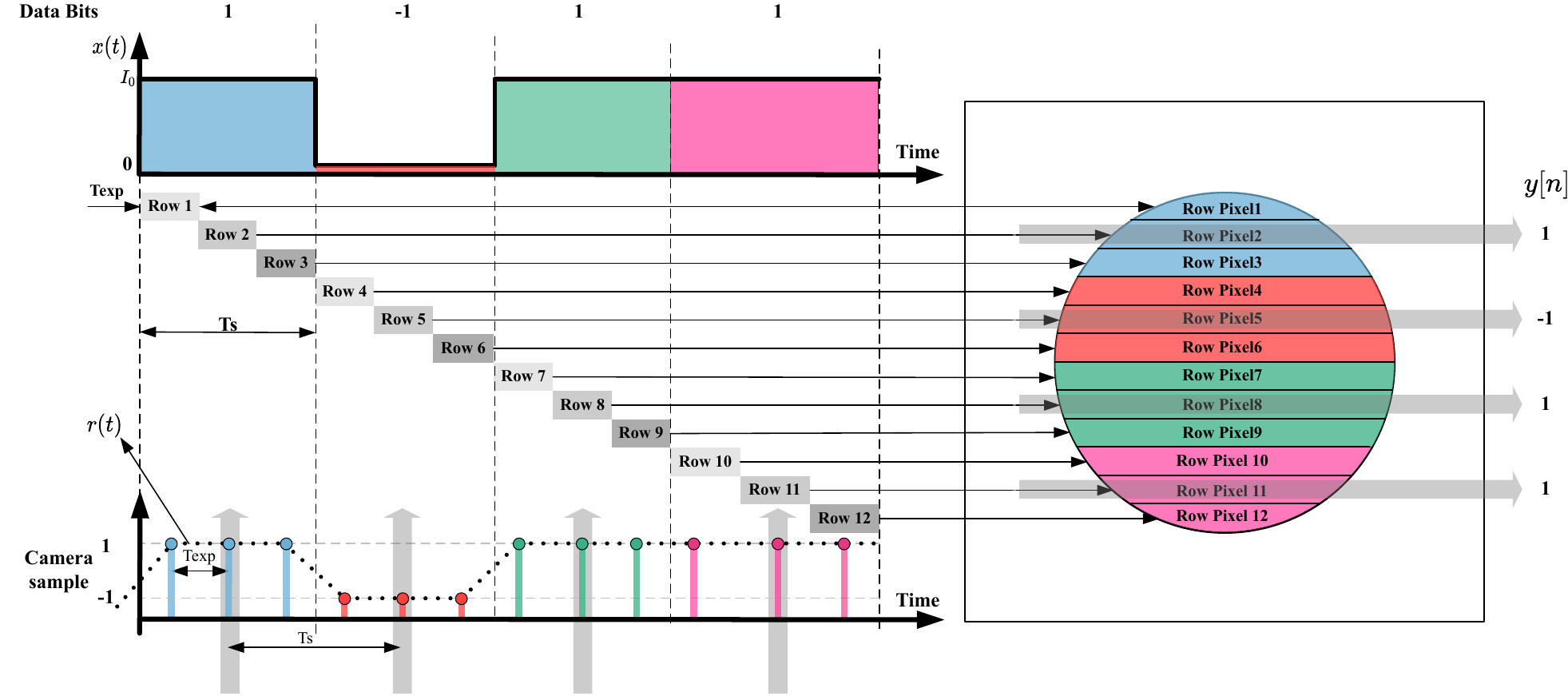}
\vspace{-4mm}
\caption{Illustration of the rolling shutter camera sensor procedures with perfect time synchronization between VLC transmitter and OCC receiver. The intensity modulated symbols with duration $T_{\rm s}$ are integrated during a time period equal to the exposure time $T_{\rm exp}$. The discrete time signal sequence that is found at the bottom is equivalent to the stripe pattern that is observed on the camera image that is found on the right-hand side part of the figure. }
\label{fig:2.2}
\vspace{-1em}
\end{figure*}



In a rolling-shutter camera, each pixel row of the captured image is exposed to the received light intensity for a time duration $T_{\rm exp}$, which is known as the exposure time. It is possible to show that the outcome of this integration process that takes place with the light intensity that reaches each pixel sensors of the rolling-shutter camera is equivalent to the one obtained when applying a convolution of the received VLC signal in each pixel with the rectangular matched filter
\begin{equation}
g_{\text{m}}(t)=
\begin{cases}
G_{\rm cam}, & 0 \leq t \leq T_\text{exp},\\[4pt]
0, & \text{otherwise},
\end{cases}
\label{eq:2.5}
\end{equation}
where $G_{\rm cam}$ is a scaling factor that is proportional to the quantum efficiency and gain of the optical camera sensor. 

Thus, the received signal after the matched-filtering processing in reception can be written as
\begin{equation}
r(t) = [s(t)*h(t)+n(t)]*g_{\rm m}(t),
\label{eq:2.6}
\end{equation}
where $h(t)$ is the effect of the optical wireless channel, $n(t)$ represents the AWGN added during the process. Finally, the continuous time-domain signal in~\eqref{eq:2.6} is then sampled by the rolling-shutter camera at its row sweep rate $f_{\text{row}}=1/T_{\text{exp}}$, producing a discrete-time sequence
\begin{align}
\label{eq:2.7}
y[n] = \sum_{k=-\infty}^{\infty} r[k]\,\delta[n-k], 
\qquad 
r[k] \triangleq r(kT_{\text{exp}}),
\end{align}
where $\delta[k]$ is the Kronecker delta function. Note that this discrete time-domain signal contains ISI introduced by the low-pass response of the LED and the lack of time synchronization between the transmitted rectangular pulses and the row sweep time modeled as a matched-filtering processing. 

To mitigate the ISI in the received signal samples, a linear equalizer with impulse response $g_{\text{eq}}[n]$ can be applied to the received sample sequence $y[n]$. Then, the post-equalized received symbol sequence can be expressed as
\begin{align}
\label{eq:2.8}
y_{\rm eq} [n] = y[n]*g_{\text{eq}}[n]=\sum_{k=-\infty}^{\infty}y[k] \, g_{\text{eq}}[n-k],
\end{align}
whose output can be fed into a slicer that will determine the most likely transmitted symbol $\widehat{a}_n$ from the the $M$-PAM alphabet based on the equalized signal sample.


\subsection{Equivalent model of the rolling shutter camera processing}
\label{sec:2b}

As explained in Section~\ref{sec:2a}, it is possible to show that the processing done at the rolling-shutter camera sensor is equivalent to performing a matched-filtering, in which the per-row exposure time acts as a shifted integration kernel that defines the impulse response $g_m(t)$ presented in~\eqref{eq:2.5}. 

This procedure is now illustrated in Fig.~\ref{fig:2.2} where, without loss of generality, the transmitted symbol sequence is \mbox{$a_n = \{1,-1,1,1\}$} and the VLC symbol duration is $T_s=3T_\text{exp}$. Under the assumption of perfect time synchronization and ideal optical wireless channel, the integration process that is performed per image row in the rolling-shutter camera converts the incident light intensity into a pixel intensity, whose value is equal to the one $r(t)$ (dotted black line) obtained by applying a rectangular matched-filter of duration $T_{\exp}$ to~$x(t)$. Since perfect time synchronization is assumed, all pixel rows from $1$ to $12$ sample the transmitted VLC waveform without any ISI. Due to that, any row pixel that is exposed to each transmitted symbol can be used to decode the received symbol sequence.

However, when there is lack of synchronization between VLC transmitter and OCC receiver, the middle row pixel of each symbol (i.e., pixel rows $2$, $5$, $8$ and $11$ of the given example) should be chosen to detect the transmitted symbol from the received signal samples. This lack of synchronization can be interpreted as a horizontal time offset in the exposure time window with respect to the VLC symbol boundaries, making edge row pixels mix adjacent symbol energy, which can be interpreted as ISI. Therefore, only the amplitude of the received signal in the middle pixel row samples remains stable, and should be used for a proper decoding of the received signal sequence. This is the main problem to be solved that limits the transmit symbol rate to have an ISI-free OCC system.

\section{Modeling and Mitigation of ISI in OCC}
\label{sec:3}

This section analyzes the synchronous and asynchronous sampling mechanisms in rolling-shutter OCC,
illustrating how timing misalignment leads to ISI and how equalization mitigates its impact to achieve the maximum symbol rate.

\subsection{Effect of Lack of Time Synchronization on OCC receiver}
\label{sec:3a}

From now on, the symbol time is set to be equal to the exposure time to maximize the symbol rate (i.e., $T_s=T_\text{exp}$).
\begin{figure}[!t]
\centering
\includegraphics[width=0.48\textwidth]{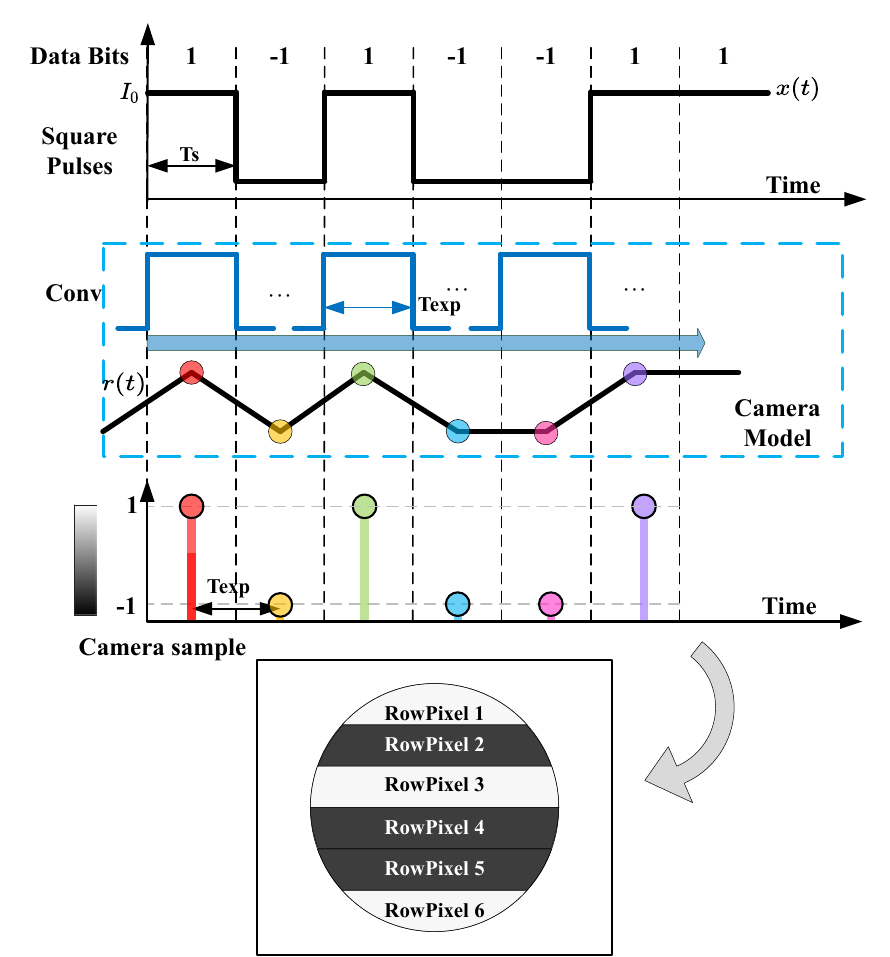} 
\vspace{-2mm}
\caption{Illustration of the output signal obtained by the equivalent OCC system model in case of perfect synchronization between transmitter and receiver. Image stripes at the bottom do not show ISI as expected in this situation.}
\vspace{-3mm}
\label{fig_3}
\end{figure}
Fig.~\ref{fig_3} illustrates this case assuming perfect synchronization when the transmitted symbol sequence is  \mbox{$a_n=\{1,-1,1,-1,-1,1\}$}. Instead of the modeling row sweep integration process explicitly, the camera operation is represented by the matched-filter $g_{\text{m}}(t)$ giving the corresponding output signal $r(t)$, from which the colored circle markers denote the discrete samples $y[n]$ at the row sweep rate $f_{\rm row}$. Since symbol time and row sweep time are perfectly synchronized, each symbol is captured by a single row. The resulting camera image clearly displays six distinct pixel rows corresponding to the six transmitted symbols, indicating perfect symbol reception for each row.

Next, we consider the case of imperfect symbol synchronization, where the symbol boundaries are misaligned with the row exposures, as illustrated in Fig.~4.
\begin{figure}[!t]
\centering
\includegraphics[width=0.48\textwidth]{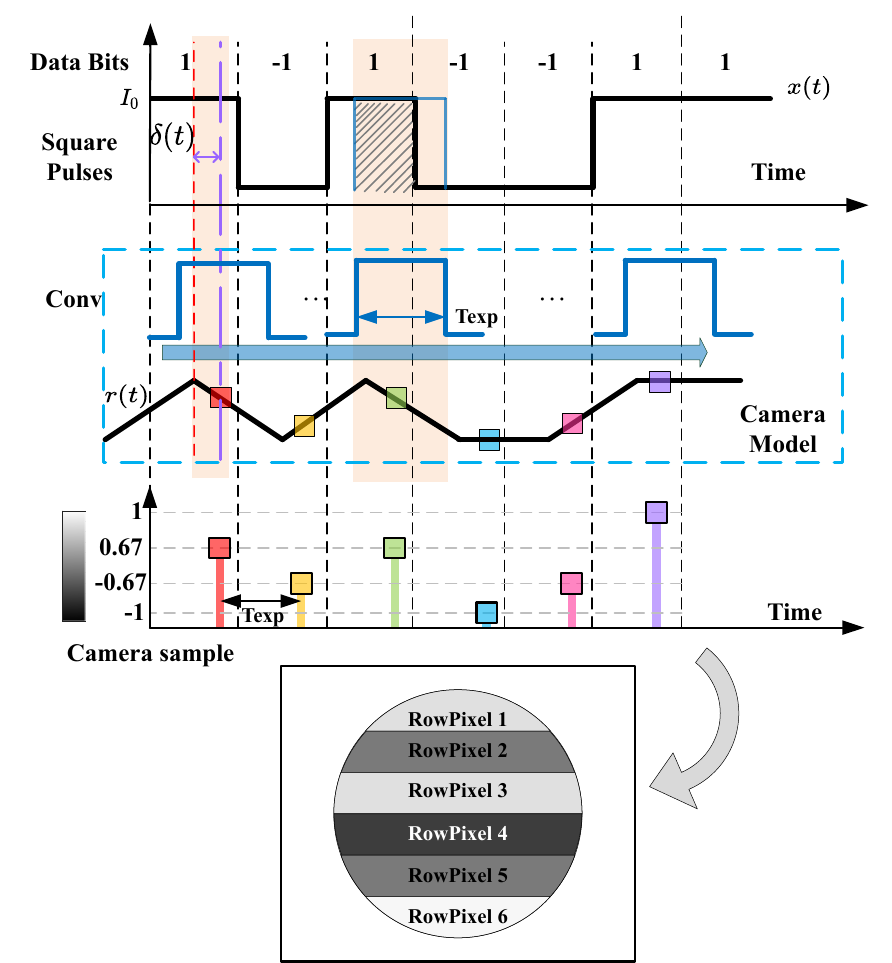} 
\vspace{-2mm}
\caption{Illustration of the output signal obtained by the equivalent OCC system model in case of a time offset $\delta(t)$ between transmitter and receiver. Image stripes at the bottom do show ISI that depends on the time offset.}
\label{fig_4}
\vspace{-3mm}
\end{figure}
A timing offset $\delta(t)$ is introduced at the camera receiver. The received signal $r(t)$ remains the convolution of the transmitted rectangular pulse with the matched filter, and its overall shape is therefore unchanged. However, the timing offset shifts the sampling instants, causing the receiver to sample slightly earlier or later than the ideal symbol boundaries. This misalignment results in ISI, as part of one symbol overlaps with adjacent samples. The sampled signal can be expressed by a three-tap ISI model:
\begin{align}
\label{eq:3.1}
y[n] = h_{0}\,a[n]+h_1\,a[n+1] + h_2\,a[n+2]
\end{align}
where coefficients $h_{0}, h_1, h_2$ represent the inter-symbol coupling coefficients, which can also be interpreted as the effective channel coefficients of the camera. As illustrated in Fig.~4, the color square markers indicate camera samples. For instance, the red sample is slightly biased toward the yellow symbol, corresponding to the coefficients $h_{0}=0.12$, $h_1=0.81$ and $h_2=0$. Moreover, when two consecutive bits are identical, only the boundary sample is affected because the constant intensity region reduces ISI. In contrast, when consecutive bits differ, both neighboring samples are affected by ISI.
This effect can also be observed in the camera image, where six row pixels exhibit mixed intensities from adjacent symbols.
\begin{figure*}
\centering
\includegraphics[width=0.95\textwidth]{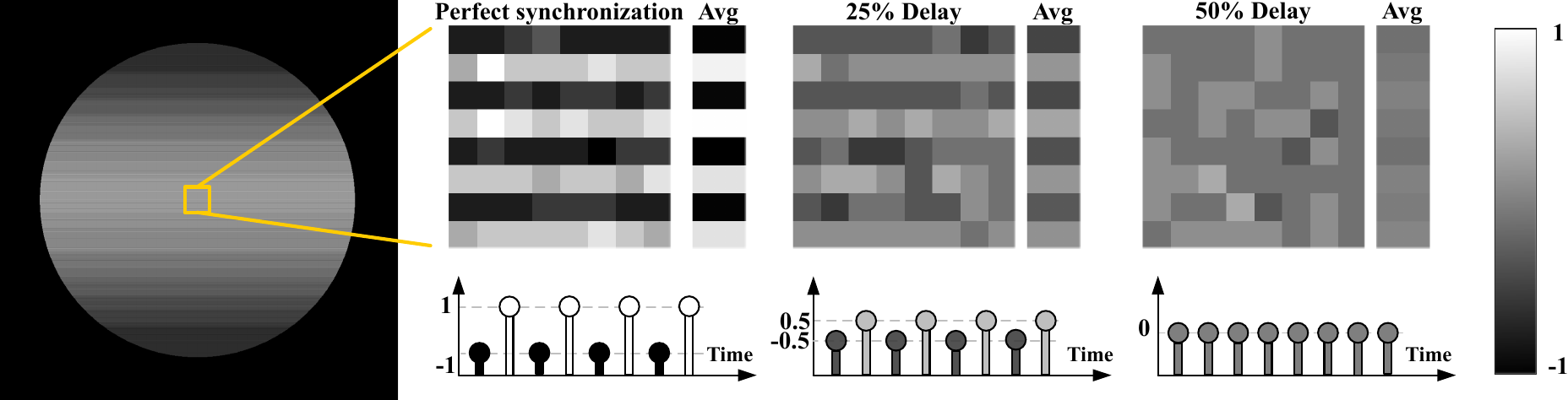} 
\vspace{-2mm}
\caption{Experimental demonstration of received signal by the rolling-shutter camera in case of different time offsets measured as percentage of symbol time.}
\label{fig_5}
\vspace{-2mm}
\end{figure*}

To further illustrate this effect, a simple experiment was conducted to capture real camera images. As shown in Fig.~5, a square-wave signal with a frequency of $f=125$\,kHz was transmitted to evaluate the impact of lack of time synchronization. Under perfect synchronization, the symbol boundaries between different symbols are clearly distinguishable after average the row pixel values. When a timing offset equal to the $25\%$ of the $T_{\rm s}$ is introduced, ISI appears and the received symbol sample levels take values closer to the central values. In the worst case, strong ISI completely blurs the symbol boundaries, making accurate symbol recovery impossible.
\subsection{Channel Estimation and  Equalizer}

In practical OCC systems, imperfect synchronization of captured images cannot be avoided because the row sweep processing  cannot synchronize to the clock of the VLC transmitter. Consequently, the timing offset $\delta(t)$ behaves as a random variable, leading each captured image to experience a distinct channel response. As a result, the channel coefficients in~\eqref{eq:3.1} should be first estimated individually for each captured image. 
In addition, the low-pass response of the LED will also distort the transmitted rectangular waveform, increasing the ISI. To recover the transmitted signal in this situation, accurate channel estimation is required. For this, a short preamble should be added in the VLC frame, i.e.,
\begin{align}
\label{eq:3.2}
 \mathbf{a_\text{frame}} = [\mathbf{a_{\text{pre}}}, \mathbf{a_\text{pld}}]^{\rm T},
\end{align}
where $\mathbf{a}_{\text{pre}} = [a_0, \ldots, a_{N_{\text{pre}}-1}]$ denotes the preamble sequence and $\mathbf{a}_{\text{pld}} = [a_{N_{\text{pre}}}, \ldots, a_{N_{\text{pre}}+N_{\text{pld}}-1}]$ represents the payload symbols, with $N_\text{pre}$ and $N_\text{pld}$ equal to the length of both parts of the frame. 
The received signal can be expressed as 
\begin{align}
\label{eq:3.3}
y[n] = \sum_{k=0}^{L_\text{c}} \, h_k \, a_{\text{frame}}[n-k] + w[n],
\end{align}
where $L_c$ is the length of channel coefficients, and the signal is modeled as AWGN samples with power  $\sigma_{\rm n}^2$. The corresponding received preambles are then given by 
\begin{align}
\label{eq:3.4}
\mathbf{y_{\text{pre}}}=[y[n_0],y[n_0+1],\ldots,y[n_0+N_{\text{pre}}-1]]^{\rm T},
\end{align}
where $n_0$ indicates the index of the first symbol of the preamble within the received signal $y[n]$. By construdcting a Toeplitz matrix $A_{\text{pre}}$ from the transmitted preamble sequence, the Least Squares~(LS) channel estimation is given by 

\begin{align}
\label{eq:15}
\hat{\mathbf{h}}_{c}
= \big(\mathbf{A}_{\mathrm{pre}}^{H}\mathbf{A}_{\mathrm{pre}}\big)^{-1}
  \mathbf{A}_{\mathrm{pre}}^{H}\,\mathbf{y}_{\mathrm{pre}}.
\end{align}
Then, when Zero Forcing~(ZF) equalization is applied,
\begin{align}
\label{eq:16}
(g_{eq}[n]*\hat{h}_{c}[n])=\delta[n-d]
\end{align}
where the $L_{\rm eq}$ is the number of taps of the ZF equalizer for the target delay $d\in\{0,\ldots,L_c+L_{eq}-2\}$, and $\delta[n]$ is the impulse response in the time domain.
To obtain the ZF equalizer coefficients, the LS solution is given by
\begin{align}
\label{eq:16}
\mathbf{g}_{\rm eq}=(\mathbf{H}^{H}\mathbf{H})^{-1}\mathbf{H}^H\mathbf{e}_d,
\end{align}
where $\mathbf{H}$ is the Toeplitz matrix built from $\mathbf{\hat{h}}_c$ and $\mathbf{e}_d$ is the canonical basis vector with a value of unit 1 at index d.
The resulting equalizer $\mathbf{g}_{\rm eq}$ is then applied to the received payload to compensate for the channel distortion and recover the estimated symbol sequence $\hat{\mathbf{a}}_n$.

\section{Simulation and Experiment results}
\label{sec:4}
The experimental setup is shown in Fig.~6. The transmitted signal is generated using GNU Radio and sent through a USRP (NI N210) software-defined radio (SDR) device. An audio Amplifier (AMP) is used to increase the signal amplitude from 1 to 2$V_{\rm pp}$ , and a DC bias is added to drive the Thorlabs LED (LIUCWHA) transmitter. On the receiver side, both a PD (PDA100A2) and a camera (Alvium 1800 U-500m) are used to capture the optical signal.
\begin{figure}[t]
\centering
\includegraphics[width=0.4\textwidth]{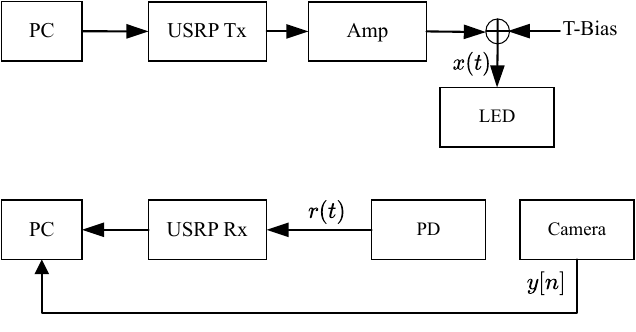} 
\vspace{-2mm}
\caption{Software-defined experiment demonstration setup including a fast-response photodetector and a rolling-shutter camera.}
\label{fig_6}
\vspace{-3mm}
\end{figure} 
\begin{figure*}
\centering
\includegraphics[width=0.98\textwidth]{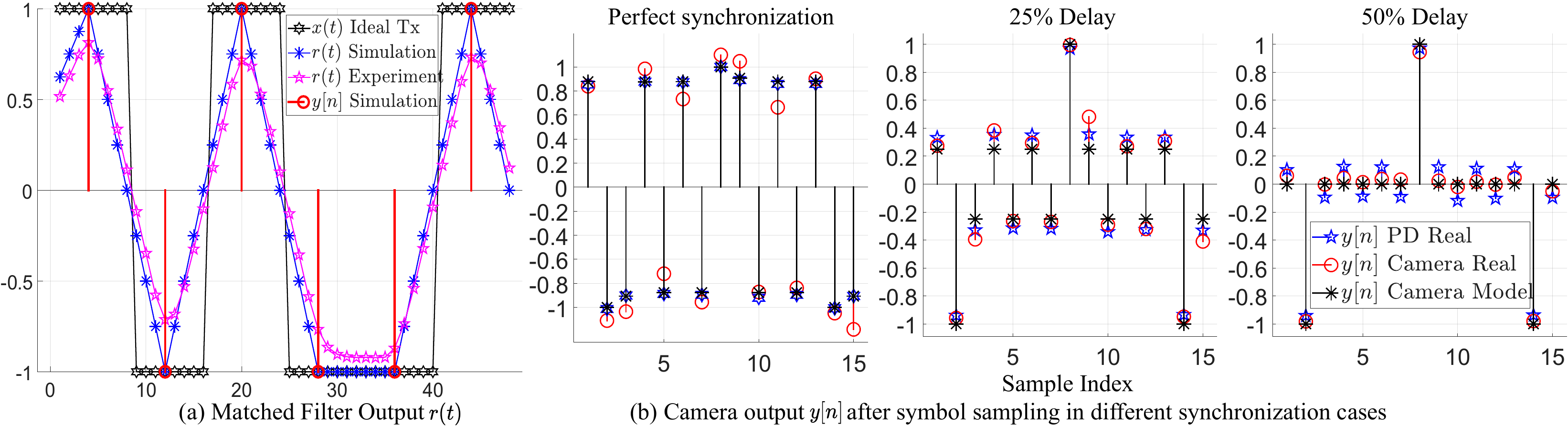} 
\vspace{-3mm}
\caption{Simulations of the equivalent OCC system model and experimental validation with rolling-shutter camera signal samples for different time offsets.}
\label{fig:7}
\vspace{-2mm}
\end{figure*}

Fig.~7(a) presents the simulation results of the proposed camera model, compared against experimental measurements for validation. The black curve with the hexagram marker represents the transmitted signal $x(t)$ with an OSF of 8. The blue curve with the star marker shows the simulated received signal $r(t)$ after the matched-filter before the symbol sampling assuming an ideal optical wireless channel. The purple curve with the pentagram marker corresponds to the experimental  waveform measured by the PD.  In this setup, the PD sampling rate is $1$~MHz, while the camera maximum row sweep rate is $125$~kHz. As a result, the PD can capture the signal $r(t)$ after the convolution for the rectangular matched-filter, which is otherwise hidden by the limited sampling rate of the camera. The PD measurements closely match the simulation results, with only minor deviations caused by the low-pass response of LED.

Fig.~7 (b) illustrates the alignment between different synchronization conditions. The black points with the star marker are the camera model simulation $y[n]$. The blue points with the pentagram marker are the real $r(t)$ from PD after symbol sampling. And the red points with the circular marker indicate the real camera output $y[n]$. Even under perfect synchronization, the camera cannot capture ideal symbol levels due to the LED low-pass response. Consecutive identical symbols exhibit slightly higher intensity than alternating ones, indicating that in practice channel equalization remains necessary even under almost perfect time synchronization between VLC transmitter and OCC receiver. In the $25\%$ of $T_{\rm s}$ time offset case, the simulation and experimental results remain closely matched. However, at a time offset equal to the $50\%$ of the $T_{\rm s}$, severe ISI appears with many samples clustering around zero amplitude, making this condition unsuitable for reliable data recovery.

 Fig.~8 shows the histogram of the received VLC symbols before (blue bars) and after (red bars) ZF equalization is applied in reception. In this latter case, the received signal samples are clearly clustered around the transmitted signal samples. This demonstrates that ZF equalization effectively mitigates the ISI. Using this method, the maximum achievable symbol rate can be realized at the camera receiver. Although only BPSK modulation was presented here, the proposed approach can be readily extended to higher modulation orders.

\begin{figure}[!t]
\centering
\includegraphics[width=0.42\textwidth]{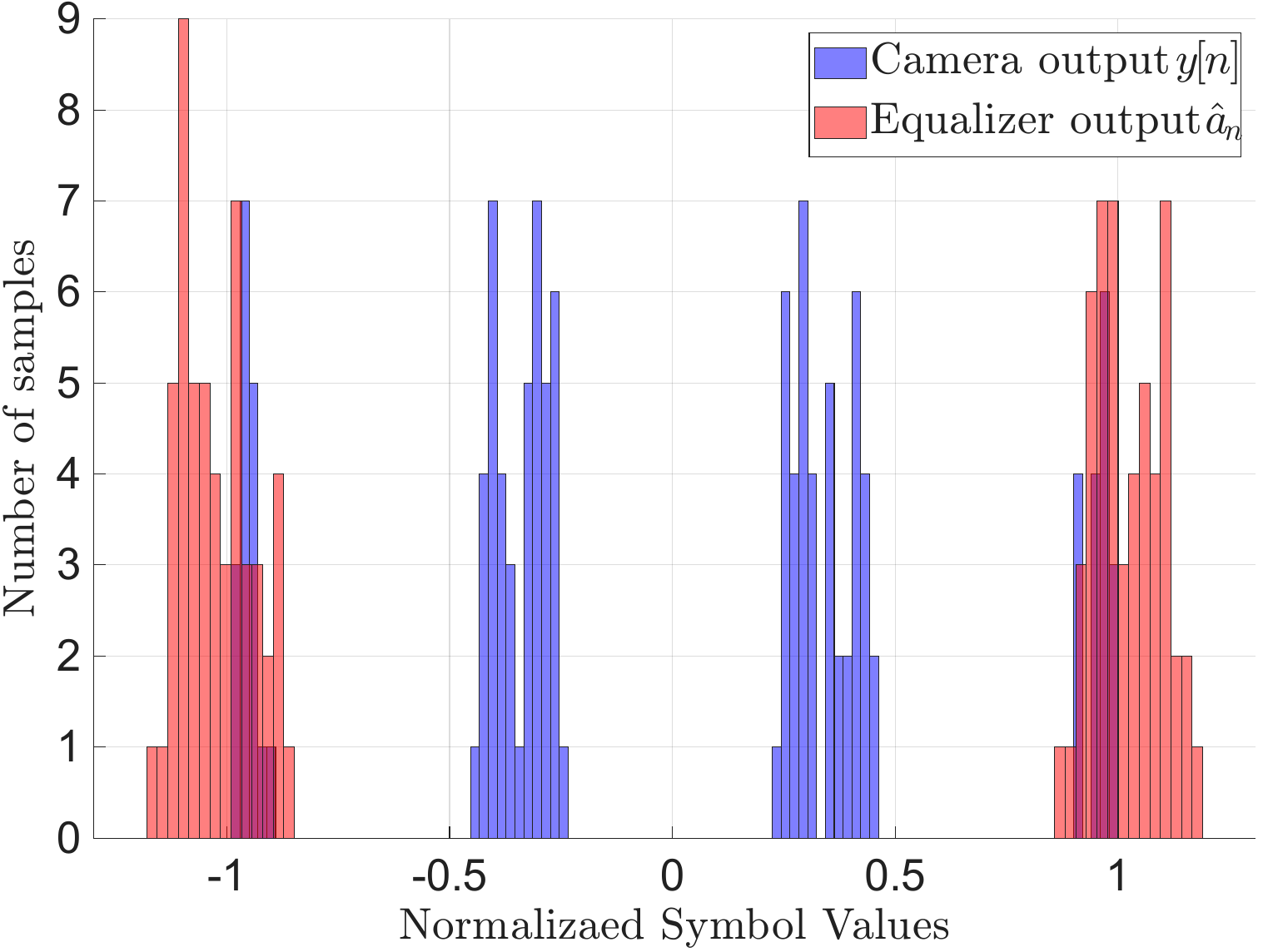} 
\vspace{-2mm}
\caption{Histogram of the signal samples received by the rolling-shutter camera for time offset $\delta(t) = 0.25 T_{\rm s}$ before (blue) and after (red) equalization.}
\label{fig_8}
\vspace{-3mm}
\end{figure}

\section{Conclusion}
\label{sec:5}

This paper demonstrated that the processing done by a rolling-shutter OCC receiver is equivalent to a matched-filtering processing with a rectangular pulse with duration~$T_{\exp}$ and symbol sampling rate equal to the row sweep rate of the rolling-shutter camera. Based on the developed model, which is well-known in digital communications, it was possible to demonstrate that the lack of synchronization between VLC transmitter and OCC receiver takes the form of ISI. By using simple linear equalization, ISI in the received OCC signal samples can be notably mitigated, offering the possibility to achieve the maximum symbol rate imposed by the rolling-shutter camera. Comprehensive simulations and experimental results validated the accuracy of the proposed model and demonstrated the effectiveness of linear equalization in achieving reliable and high-rate OCC signal receptions.
\bibliographystyle{IEEEtran}
\bibliography{References}

\end{document}